\documentclass[aps,prl,twocolumn,superscriptaddress,showpacs,letterpaper]{revtex4}

\usepackage{color}
\usepackage{bm}
\usepackage{dcolumn}
\usepackage{graphicx}
\usepackage{subfigure}

\newcommand{\hept}{$^4$He$(\vec{e},e^\prime\vec{p}\,)\,^3$H }
\newcommand{\hep}{$^1$H$(\vec{e},e^\prime\vec{p}\,)$ }

\begin{document}

\title{Polarization Transfer in the $\bm{^4}$He$\bm{(\vec{e},e^\prime\vec{p}\,)\,^3}$H Reaction at $\bm{Q^2 = 0.8}$ and $\bm{1.3}$ (GeV$/c$)$\bm{^2}$}

\newcommand*{\CNU}{Christopher Newport University, Newport News, Virginia 23606}
\newcommand*{\RUTGERS}{Rutgers, The State University of New Jersey, Piscataway, New Jersey 08854}
\newcommand*{\CSU}{California State University, Los Angeles, Los Angeles, California 90032}
\newcommand*{\DU}{Dalhousie University, Halifax, Nova Scotia, Canada}
\newcommand*{\DUKE}{Duke University, Durham, North Carolina 27708}
\newcommand*{\MIT}{Massachusetts Institute of Technology, Cambridge, Massachusetts 02139}
\newcommand*{\KENT}{Kent State University, Kent, Ohio 44242}
\newcommand*{\WM}{College of William and Mary, Williamsburg, Virginia 23187}
\newcommand*{\ODU}{Old Dominion University, Norfolk, Virginia 23529}
\newcommand*{\ROME}{INFN, Sezione Sanit\'a and Istituto Superiore di Sanit\'a, Laboratorio di Fisica, I-00161 Rome, Italy}
\newcommand*{\UVA}{University of Virginia, Charlottesville, Virginia 22904}
\newcommand*{\KIPT}{Kharkov Institute of Physics and Technology, Kharkov 310108, Ukraine}
\newcommand*{\NSU}{Norfolk State University, Norfolk, Virginia 23504}
\newcommand*{\SMU}{Saint Mary's University, Halifax, Nova Scotia, Canada}
\newcommand*{\SNU}{Seoul National University, Seoul, Korea}
\newcommand*{\TELAVIV}{Tel Aviv University, Tel Aviv 69978, Israel}
\newcommand*{\UCM}{Universidad Complutense de Madrid, E-28040 Madrid, Spain}
\newcommand*{\ARGONNE}{Argonne National Laboratory, Argonne, Illinois}
\newcommand*{\HU}{Hampton University, Hampton, Virginia 23668}
\newcommand*{\SCAROLINA}{University of South Carolina, Columbia, South Carolina 29208}
\newcommand*{\JLAB}{Thomas Jefferson National Accelerator Facility, Newport News, Virginia 23606}
\newcommand*{\GWU}{The George Washington University, Washington, DC 20052}
\newcommand*{\WEIZ}{Weizmann Institute of Science, Rehovot 76100, Israel}
\newcommand*{\IEM}{Instituto de Estructura de la Materia, CSIC, E-28006 Madrid, Spain}

\author {M.~Paolone} \affiliation{\SCAROLINA}
\author {S.P.~Malace} \affiliation{\SCAROLINA}
\author {S.~Strauch} \affiliation{\SCAROLINA}
\author {I.~Albayrak} \affiliation{\HU}
\author {J.~Arrington} \affiliation{\ARGONNE} 
\author {B.L.~Berman} \affiliation{\GWU}
\author {E.J.~Brash} \affiliation{\CNU} 
\author {B.~Briscoe} \affiliation{\GWU}
\author {A.~Camsonne} \affiliation{\JLAB}
\author {J.-P.~Chen} \affiliation{\JLAB}
\author {M.E.~Christy} \affiliation{\HU}
\author {E.~Chudakov} \affiliation{\JLAB}
\author {E.~Cisbani} \affiliation{\ROME}
\author {B.~Craver} \affiliation{\UVA}
\author {F.~Cusanno} \affiliation{\ROME}
\author {R.~Ent} \affiliation{\JLAB}
\author {F.~Garibaldi} \affiliation{\ROME}
\author {R.~Gilman} \affiliation{\RUTGERS} \affiliation{\JLAB}
\author {O.~Glamazdin} \affiliation{\KIPT}
\author {J.~Glister} \affiliation{\SMU} \affiliation{\DU}
\author {D.W.~Higinbotham} \affiliation{\JLAB}
\author {C.E.~Hyde-Wright} \affiliation{\ODU}
\author {Y.~Ilieva} \affiliation{\GWU}
\author {C.W.~de~Jager} \affiliation{\JLAB}
\author {X.~Jiang} \affiliation{\RUTGERS}
\author {M.K.~Jones} \affiliation{\JLAB}
\author {C.E.~Keppel} \affiliation{\HU} 
\author {E.~Khrosinkova} \affiliation{\KENT}
\author {E.~Kuchina} \affiliation{\RUTGERS}
\author {G.~Kumbartzki} \affiliation{\RUTGERS}
\author {B.~Lee} \affiliation{\SNU}
\author {R.~Lindgren} \affiliation{\UVA} 
\author {D.J.~Margaziotis} \affiliation{\CSU}
\author {D.~Meekins} \affiliation{\JLAB}
\author {R.~Michaels} \affiliation{\JLAB}
\author {K.~Park} \affiliation{\JLAB}
\author {L.~Pentchev} \affiliation{\WM}
\author {C.F.~Perdrisat} \affiliation{\WM} 
\author {E.~Piasetzky} \affiliation{\TELAVIV}
\author {V.A.~Punjabi} \affiliation{\NSU} 
\author {A.J.R.~Puckett} \affiliation{\MIT}
\author {X.~Qian} \affiliation{\DUKE}
\author {Y.~Qiang} \affiliation{\MIT}
\author {R.D.~Ransome} \affiliation{\RUTGERS}
\author {A.~Saha} \affiliation{\JLAB}
\author {A.J.~Sarty} \affiliation{\SMU}
\author {E.~Schulte} \affiliation{\RUTGERS}
\author {P.~Solvignon} \affiliation{\ARGONNE} 
\author {R.R.~Subedi} \affiliation{\KENT} 
\author {L.~Tang} \affiliation{\HU} 
\author {D.~Tedeschi} \affiliation{\SCAROLINA}
\author {V.~Tvaskis} \affiliation{\HU}
\author {J.M.~Udias} \affiliation{\UCM}
\author {P.E.~Ulmer} \affiliation{\ODU} 
\author {J.R.~Vignote} \affiliation{\IEM}
\author {F.R.~Wesselmann} \affiliation{\NSU}
\author {B.~Wojtsekhowski} \affiliation{\JLAB}
\author {X.~Zhan} \affiliation{\MIT}

\collaboration{The E03-104 Collaboration}
     \noaffiliation

\date{\today}

\begin{abstract}
Proton recoil polarization was measured in the quasielastic \hept reaction at $Q^2$ = 0.8 (GeV$/c$)$^2$ and 1.3 (GeV$/c$)$^2$ with unprecedented precision.  The polarization-transfer coefficients are found to differ from those of the \hep reaction, contradicting a relativistic distorted-wave approximation, and favoring either the inclusion of medium-modified proton form factors predicted by the quark-meson coupling model or a spin-dependent charge-exchange final-state interaction. For the first time, the polarization-transfer ratio is studied as a function of the virtuality of the proton. \end{abstract}


\maketitle

\begin{acknowledgments}
\end{acknowledgments}

	Electron-nucleon scattering is a powerful tool for probing the structure of nucleons. For over a decade, access to high-quality polarized electron beams has allowed the nucleon's electromagnetic properties to be explored through measurement of polarization observables. In elastic electron-nucleon scattering, the polarization-transfer technique allows measurement of the Sachs form-factor ratio $G_E/G_M$, that is directly proportional to the ratio of transverse and longitudinal polarization observables $P_x'/P_z'$ in the single-photon exchange approximation \cite{Akhiezer:1974em, coordsys}. This technique \cite{Perdrisat:2006hj} benefits from a large cancellation of systematic uncertainties, unlike the Rosenbluth separation technique, which relies on repeated cross-section measurements. Several recent experiments have extracted $G_E/G_M$ of the proton using this method \cite{Punjabi:2005wq, Gayou:2001qt, Crawford:2006rz, Ron:2007vr}.

	The question of if and how the nucleon structure is modified within the nuclear medium has been hotly debated since the discovery of the nuclear EMC effect, which showed that quark momentum distributions within nuclei differ from those within free nucleons. Indeed, a deviation of $G_E$ and $G_M$ of a nucleon immersed in a nuclear medium from their free-space values is predicted by Lu \textit{et al.} \cite{Lu:1997mu, Saito:2005rv} using the quark-meson coupling (QMC) model. These results are consistent with experimental constraints from the Coulomb Sum Rule; see \cite{Meziani:1992xr,E01-015}. In addition to the QMC model, many other model calculations predict the in-medium modification of nucleon structure; for recent examples see \cite{Frank:1995pv, Yakhshiev:2002sr, Smith:2004dn, Horikawa:2005dh}. Ciofi degli Atti \textit{et al.} predict that the proton form factors are strongly correlated with the excitation of the residual system and the virtuality of the ejected proton \cite{CiofidegliAtti:2007vx}.

	The polarization-transfer technique can be used to help settle this question using quasi-elastic nucleon knockout. In that case, the ratio $G_E/G_M$ remains approximately proportional to $P_x'/P_z'$, allowing modifications of the form factors to be determined. However, in-medium nucleon interactions complicate this picture, and even raise the question as to whether the concept of medium modifications is a meaningful one, due to the complex nature of the in-medium interaction. Predictions from Schiavilla \cite{Schiavilla:2004xa} contend that final-state interactions (FSIs) including charge exchange processes and meson exchange currents lead to a quenching of 10\% in the polarization- transfer ratio $P_x'/P_z'$ in the quasielastic scattering reaction \hept compared with the free-space reaction $^1$H$(\vec{e},e^\prime\vec{p}\,)$. The correct treatment of FSIs in a model calculation is essential to separate any unconventional medium effects from FSIs, since both influence the polarization-transfer observables. To help settle this debate precision measurements are needed with the polarization-transfer coefficients $P_x'/P_z'$ mapped in detail in a region of low ($<$ 100 MeV/$c$) missing momentum, where such FSI complications are minimized, and as a function of the virtuality of the ejected proton. Dependence on the latter is a simple and straightforward corollary of models with medium modifications.

	This Letter reports on measurements of the polarization-transfer coefficients $P_x'$ and $P_z'$ in the quasi-elastic \hept reaction preformed at Jefferson Lab in Hall A: experiment E03-104.  Data were taken at four-momentum transfers of $Q^2$ = 0.8 and 1.3 (GeV$/c$)$^2$ within a missing-momentum range $<$ 160 MeV$/c$. The $^4$He target was chosen for its high nuclear density and relative theoretical modeling simplicity.  A recent study of the EMC effect \cite{Seely:2009gt} has shown that the effect on nucleons in $^4$He is comparable to the effect on nucleons in $^{12}$C. The low missing-momentum regime was chosen to reduce the contribution from many-body effects, although a weaker contribution from in-medium modification effects is expected. Additional \hep scattering data also were taken to provide unmodified proton scattering measurements as a basis for comparison.  The carbon analyzing power of the polarimeter was also extracted from the \hep data.

	Kinematic settings for the present experiment are given in Table \ref{tab:kin}.  For both \hep and $^4$He$(\vec{e},e^\prime\vec{p}\,)\,^3$H, the scattered electron and ejected proton were detected in coincidence in two high-resolution spectrometer arms. For the nine $^1$H settings, the central momenta for the proton were adjusted in $2\%$ increments from $- 8\%$ to $+ 8\%$ in order to produce similar coverage of the focal plane, as in \hept scattering. This allowes for detailed studies of the spin transport and other instrumental effects.  Beam currents up to 80 $\mu$A and beam polarizations of 85$\%$ were used.  The proton spectrometer was equipped with a focal plane polarimeter (FPP) which measures the asymmetry of polarized protons scattered from a carbon analyzer \cite{Punjabi:2005wq}.  The spin precession of the proton in the magnetic field of the spectrometer was calculated using the COSY software \cite{Berz:1997zz}. A maximum likelihood method was then employed in conjunction with the beam helicity, the carbon analyzing power, and the proton spin precession to extract the polarization of the ejected proton at the target \cite{Besset:1979sh}. The large amount of statistics accumulated in this experiment have allowed the extraction of $\mu G_E/G_M$ from the data with strict missing-energy and missing-momentum cuts to prevent any effects from diluting the polarization observables. For \hept scattering, tight cuts on the reconstructed missing mass spectrum were used to ensure that quasi-elastic knockout of the proton leaves the undetected $^3$H intact. Radiative effects due to single-photon emission \cite{PhysRevD.65.013006}, as well as radiative corrections from two-photon exchange to the polarization ratio $P_x'/P_z'$ \cite{PhysRevC.72.034612}, are predicted to be less than $0.5\%$. Radiative effects on the ratio were minimized with missing-energy and missing-momentum cuts, but no specific radiation corrections were applied to the data.

\begin{figure*}[]
\centering
\mbox{\subfigure{\includegraphics[width=1.0\columnwidth]{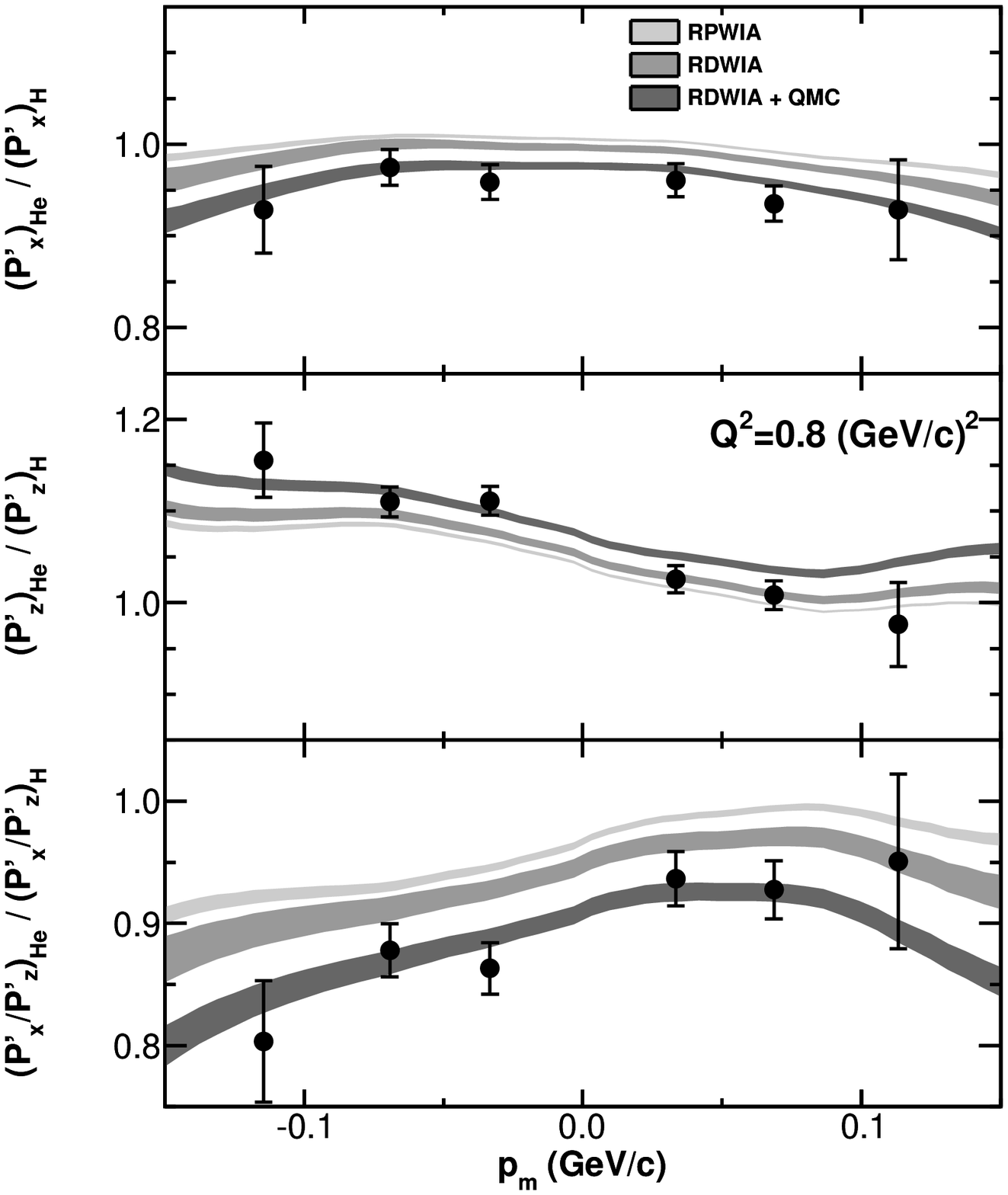}}\quad \quad
\subfigure{\includegraphics[width=1.0\columnwidth]{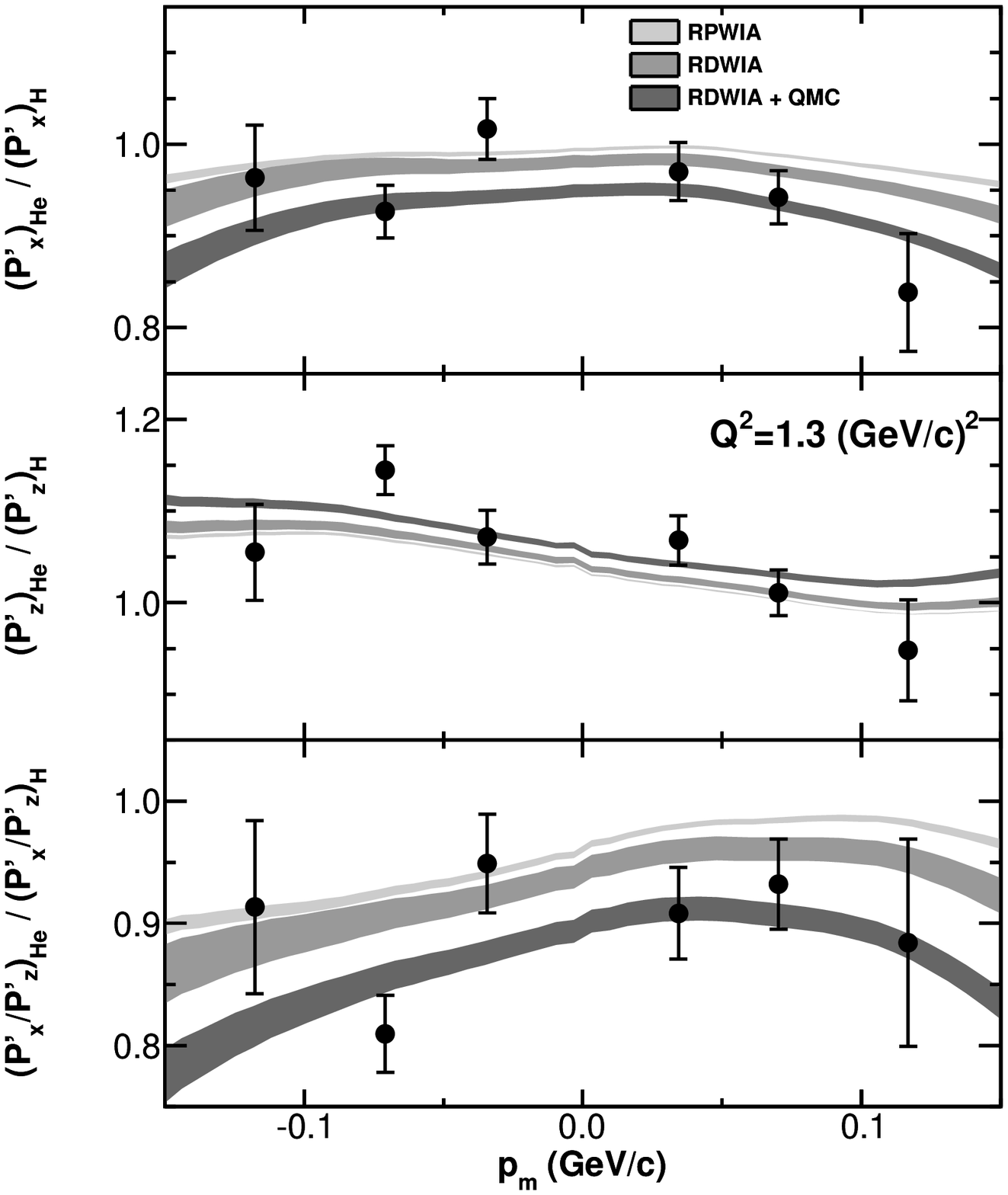} }}
\caption{The individual polarization-transfer coefficients from $^4$He normalized to $^1$H, $(P_x')_\mathrm{He}/(P_x')_\mathrm{H}$ and $(P_z')_\mathrm{He}/(P_z')_\mathrm{H}$, and the double-ratio  $R$ versus the missing momentum $p_m$ for $Q^2$= 0.8 (GeV$/c$)$^2$ (left) and $Q^2$= 1.3 (GeV$/c$)$^2$ (right). The bands represent RPWIA (light grey), RDWIA calculations (medium grey), and RDWIA + QMC calculations (dark grey) \cite{Florizone:1999ba}.  See the text for a description of the models.
} \label{fig:prec}
\end{figure*}

	Figure \ref{fig:prec} shows our results for the polarization-transfer coefficients as a function of the missing momentum.  Here, the sign of the missing momentum is positive if the component of the missing-momentum vector along the momentum-transfer direction is positive.  The individual polarization-transfer coefficients from the \hept normalized to the \hep reaction, $(P_x')_\mathrm{He}/(P_x')_\mathrm{H}$ and $(P_z')_\mathrm{He}/(P_z')_\mathrm{H}$, and the double ratio $R$ is shown along with acceptance-corrected calculations from the Madrid group \cite{Udias:1999tm, Udias:2000ig}. Here, $R$ is defined as:
\begin{equation}
R = \frac{(P_x'/P_z')_{^4\rm He}}{(P_x'/P_z')_{^1\rm H}} .
\label{eq:rexp}
\end{equation}
	The Madrid group calculations use a relativistic wave function for the bound state that reproduces the exclusive $^4$He$(e,e'p)$ cross section data \cite{Florizone:1999ba}.  The calculations are represented through bands whose variation in width depends on the nuclear current operators, $cc1$ and $cc2$ \cite{DeForest:1983vc}, and the optical potential models, MRW \cite{McNeil:1983yi} and RLF \cite{Horowitz:1985tw}, used. 
The light, medium, and dark grey bands represent calculations from a relativistic plane-wave impulse-approximation (RPWIA), relativistic distorted-wave impulse-approximation (RDWIA), and a RDWIA that includes an in-medium modified form factor as predicted by Lu \textit{et al.\,}with the QMC model \cite{Lu:1997mu}, respectively.  At both $Q^2$ = 0.8 (GeV$/c$)$^2$ and 1.3 (GeV$/c$)$^2$ the RPWIA  and RDWIA calculations overestimate the data significantly.  With RDWIA + QMC, the calculation is in better agreement with the data.  Uncertainties from model wave functions, current operators, or choice of MRW or RLF optical potentials are small which allows discrimination between the data and the conventional RDWIA calculations. The RDWIA calculations with medium-modified nucleon form factors predict a greater divergence from standard RDWIA calculations at missing momenta further from zero.  

	The expected effect on the hydrogen-normalized polarization coefficients from in-medium modified form factors can be estimated by comparing the $\vec{e}p$ elastic scattering to the quasielastic case. In elastic scattering, the polarization coefficients themselves can be expressed directly as functions of $P_x'/P_z'$. One would expect a decrease for $(P_x')_\mathrm{He}/(P_x')_\mathrm{H}$ and an increase for $(P_z')_\mathrm{He}/(P_z')_\mathrm{H}$, consistent with the overall observed quenching of R, which is indeed consistent with our data for both observables. These results are also in agreement with the full model, RDWIA + QMC.

\begin{figure}[]
\includegraphics[width=0.95\columnwidth]{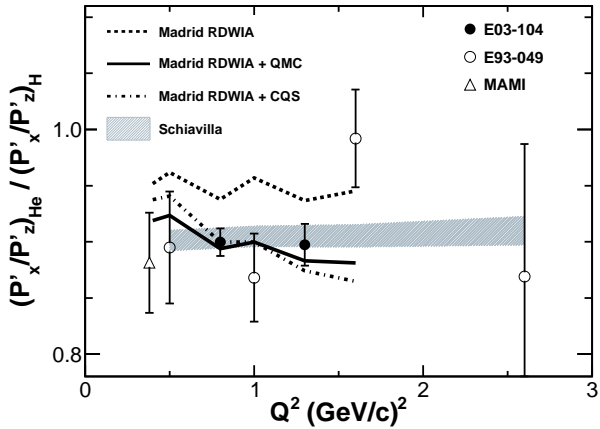}
\caption{\label{fig:rq2} Experimental results for $R$ versus $Q^2$ for E03-104 (black circles), E93-049 (open circles) \cite{Strauch:2002wu} and MAMI (open triangle) \cite{Dieterich:2000mu}.    The curves represent RDWIA (dashed), RDWIA + QMC (solid), and RDWIA + CQS (dash-dot) calculations with the current operator $cc2$ and the MRW optical potential \cite{Florizone:1999ba}.  The grey band represents Schiavilla's model \cite{Schiavilla:2004xa}; See text for details.}
\end{figure}

	In Figure \ref{fig:rq2}, results are shown as the polarization-transfer double-ratio $R$ plotted versus $Q^2$.  The results agree with previous results \cite{nopwia} from Mainz \cite{Dieterich:2000mu} and JLab experiment E93-049 \cite{Strauch:2002wu} establishing the quenching of $R$ and its $Q^2$ dependence with previously unattained confidence; additionally, the calculated $\mu G_E/G_M$ values for \hep are in good agreement with world data \cite{Punjabi:2005wq, Gayou:2001qt, Crawford:2006rz, Ron:2007vr}.  The experimental results for $R$ and $\mu G_E/G_M$ are also listed in Table \ref{tab:results}.  With data for \hep and \hept obtained under near identical experimental conditions, calculating the double-ratio $R$ results in a significant cancellation of systematic uncertainties.

	The theoretical calculations shown in Figure \ref{fig:rq2} include a RDWIA calculation with free-space proton form factors (dashed line), and RDWIA calculations that include an in-medium modified form factor as predicted by Lu \textit{et al.\,}with the QMC model \cite{Lu:1997mu} (solid line) and an in-medium modified form factor as predicted in the chiral quark soliton model by Smith and Miller \cite{Smith:2004dn} (dash-dot line).  Theoretical calculations from Schiavilla \cite{Schiavilla:2004xa} are included in Figure \ref{fig:rq2} as a grey band, and assume a missing momentum close to zero and have not been acceptance corrected.  Schiavilla shows with conventional many-body calculations that a model with free-space nucleon form factors can describe $R$ as a function of $Q^2$.  The difference in modeling the FSIs account for most of the discrepancy between Schiavilla's and the Madrid group's calculations. Schiavilla's calculation includes MEC effects paired with tensor correlations that suppress $R$ by 4\% and include both a spin-dependent and a spin-independent charge-exchange term in the final-state interaction that suppress $R$ by an additional 6\%, all of which are not included in the Madrid group's calculations.  The spin-orbit terms in Schiavilla's FSI calculations are not well constrained, and the Monte Carlo technique employed in the model calculation introduces a statistical uncertainty represented in the width of the grey band in Figure \ref{fig:rq2}.

\begin{figure}[]
\includegraphics[width=0.95\columnwidth]{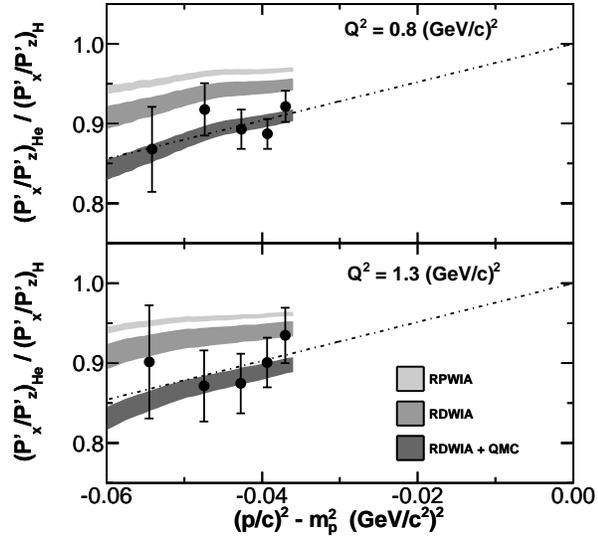}
\caption{\label{fig:virt} The double ratio $R$ versus the proton virtuality for $Q^2$= 0.8 and 1.3 (GeV$/c$)$^2$. The dashed line is a linear fit to the data constrained to have a $y$ intercept value of one at zero virtuality. The bands represent RPWIA (light grey), RDWIA calculations (grey), and RDWIA + QMC calculations (dark grey)\cite{Florizone:1999ba}.  See the text for a description of the models.}
\end{figure}

	Figure \ref{fig:virt} shows $R$ as a function of the proton virtuality, $v = p^2 - m_{\mathrm{p}}^2$.  Here, $p$ is the proton four-momentum in the $^4$He nucleus and is defined as $p^2 = (m_{\mathrm{He}} - E_{\mathrm{t}})^2 - p_{\mathrm{t}}^2$ in the impulse approximation, where $E_{\mathrm{t}}$ and $p_{\mathrm{t}}$ are respectively the energy and momentum of the undetected triton.  The dashed line is a linear fit to the data assuming $R=1$ at $v=0$, and is included as a simple approximation of the expected trend in virtuality.  The RDWIA models including medium modified proton form factors describe the data best.  The Madrid group RDWIA + QMC calculations diverge from the conventional RDWIA calculations as the virtuality moves further from zero.  Calculations from Schiavilla are not available as a function of the missing momentum or the virtuality.

	In summary, we have measured recoil polarization in the \hept reaction at $Q^2$ values of 0.8 (GeV$/c$)$^2$ and 1.3 (GeV$/c$)$^2$.  The data agree well with previously reported measurements from Mainz \cite{Dieterich:2000mu} and JLab \cite{Strauch:2002wu}, but the increased precision challenges state-of-the-art nuclear physics calculations, both with and without medium modifications. Our data allow one to study the dependence of polarization-transfer ratios as functions of missing momentum and, for the first time, proton virtuality. The data are in excellent agreement with model calculations including the medium modification of the proton form factors through the quark-meson coupling model presented by Lu \textit{et al} \cite{Lu:1997mu}, and with a chiral quark soliton model by Smith and Miller \cite{Smith:2004dn}. A model calculation by Schiavilla \cite{Schiavilla:2004xa}, which uses conventional free-space nucleon form factors, but employs a different treatment of in-medium nucleon interactions, including charge exchange processes, also agrees with the overall reduction of the polarization-transfer ratios, albeit within large uncertainties. Combining these data with similar precision induced-polarization data, directly sensitive to the amount of in-medium nucleon interactions, may lead to a definite statement in favor or against the effective use of proton medium modifications.

	The collaboration wishes to acknowledge the Hall A technical staff and the Jefferson Lab Accelerator Division for their terrific support. This work was supported by the U.S.\ Department of Energy and the U.S.\ National Science Foundation. Jefferson Science Associates operates the Thomas Jefferson National Accelerator Facility under DOE contract DE-AC05-06OR23177.

\begin{table*}[]
\centering
\caption[Table of kinematic settings for Experiment E03-104.]{Table of kinematic settings for Experiment E03-104.  Here $E_0$ is the incident beam energy, $p_{p}$ is the central momentum setting of the proton spectrometer, $\theta_{p}$ is the central angle setting for the proton spectrometer, $p_{e}$ is the central momentum setting of the electron spectrometer, and $\theta_{e}$ is the central angle setting for the electron spectrometer.}
\begin{tabular}[c]{cccccccc}
\hline
\hline
	Kinematic & $Q^2$ &  $E_0$ & & $p_{p}$ &  $\theta_{p}$ & $p_{e}$ & $\theta_{e}$ \\
	Setting &  (GeV/$c$)$^2$ & (GeV) & Target  &  (GeV/$c$) & (deg) &  (GeV/$c$) &  (deg) \\
	\hline
	A1--9 & 0.8 & 1.987 & $^1$H & 0.991$\, \pm \,8\%$ \,&  50.668 & 1.561 & $-$29.440 \\
	A10 & 0.8 & 1.987 &  $^4$He & 1.004 & 49.115 & 1.532 & $-$29.730 \\
	B1--9 & 1.3 & 2.637 & $^1$H & 1.334$\, \pm \,8\%$ & 45.289 & 1.944 & $-$29.221 \\
	B10 & 1.3 & 2.637 & $^4$He & 1.353 & 43.920 & 1.909 & $-$29.462 \\
	\hline
	\hline
\end{tabular}
	\label{tab:kin}
\end{table*}

\begin{table*}[]
\centering
\caption[Values for the polarization-transfer coefficients $P_x'$ and $P_z'$ of the ejected proton from the listed target at both four-momentum transfer settings.]{Values for the polarization-transfer coefficients $P_x'$ and $P_z'$ of the ejected proton from the listed target at both four-momentum transfer settings.  Uncertainties are listed as statistical then systematic.  Systematic uncertainties in the ratios $(P_x')_{\mathrm{He}}/(P_x')_{\mathrm{H}}$, $(P_z')_{\mathrm{He}}/(P_z')_{\mathrm{H}}$, and the double ratio $R$ mostly cancel, providing a systematic precision better than $5.0\times10^{-4}$.}
\begin{tabular}[c]{ccccc}
	\hline
	\hline
	$Q^2$ (GeV$/c$)$^2$ & $(P_x')_{\mathrm{He}}/(P_x')_{\mathrm{H}}$ & $(P_z')_{\mathrm{He}}/(P_z')_{\mathrm{H}}$ & $\mu G_E/G_M$ & $R$\\
	\hline
	0.8 & \, 1.062 $\pm$ 0.009 \,& \, 0.956 $\pm$ 0.010 \,& \, 0.901 $\pm$ 0.007 $\pm$ 0.010 \,& 0.900 $\pm$ 0.012\\
	1.3 & \, 1.064 $\pm$ 0.014 \,& \, 0.954 $\pm$ 0.015 \,& \, 0.858 $\pm$ 0.008 $\pm$ 0.019 \,& 0.897 $\pm$ 0.019\\
	\hline
	\hline
\end{tabular}
	\label{tab:results}
\end{table*}

\bibliography{e03104_hd_f1}

\begin{thebibliography}{31}
\expandafter\ifx\csname natexlab\endcsname\relax\def\natexlab#1{#1}\fi
\expandafter\ifx\csname bibnamefont\endcsname\relax
  \def\bibnamefont#1{#1}\fi
\expandafter\ifx\csname bibfnamefont\endcsname\relax
  \def\bibfnamefont#1{#1}\fi
\expandafter\ifx\csname citenamefont\endcsname\relax
  \def\citenamefont#1{#1}\fi
\expandafter\ifx\csname url\endcsname\relax
  \def\url#1{\texttt{#1}}\fi
\expandafter\ifx\csname urlprefix\endcsname\relax\def\urlprefix{URL }\fi
\providecommand{\bibinfo}[2]{#2}
\providecommand{\eprint}[2][]{\url{#2}}

\bibitem[{\citenamefont{Akhiezer and Rekalo}(1974)}]{Akhiezer:1974em}
\bibinfo{author}{\bibfnamefont{A.~I.} \bibnamefont{Akhiezer}} \bibnamefont{and}
  \bibinfo{author}{\bibfnamefont{M.~P.} \bibnamefont{Rekalo}},
  \bibinfo{journal}{Sov. J. Part. Nucl.} \textbf{\bibinfo{volume}{4}},
  \bibinfo{pages}{277} (\bibinfo{year}{1974}).

\bibitem[{coo()}]{coordsys}
\bibinfo{howpublished}{With the initial and final electron energy given as
  $\vec{k}_i$ and $\vec{k}_f$, the coordinate system is given by $\hat{z} =
  (\vec{k}_i - \vec{k}_f)/|\vec{k}_i - \vec{k}_f|$, $\hat{y} = (\vec{k}_i
  \times \vec{k}_f) / |\vec{k}_i \times \vec{k}_f|$, and $\hat{x}=\hat{y}
  \times \hat{z}$.}

\bibitem[{\citenamefont{Perdrisat et~al.}(2007)\citenamefont{Perdrisat,
  Punjabi, and Vanderhaeghen}}]{Perdrisat:2006hj}
\bibinfo{author}{\bibfnamefont{C.~F.} \bibnamefont{Perdrisat}},
  \bibinfo{author}{\bibfnamefont{V.}~\bibnamefont{Punjabi}}, \bibnamefont{and}
  \bibinfo{author}{\bibfnamefont{M.}~\bibnamefont{Vanderhaeghen}},
  \bibinfo{journal}{Prog. Part. Nucl. Phys.} \textbf{\bibinfo{volume}{59}},
  \bibinfo{pages}{694} (\bibinfo{year}{2007}).

\bibitem[{\citenamefont{Punjabi et~al.}(2005)}]{Punjabi:2005wq}
\bibinfo{author}{\bibfnamefont{V.}~\bibnamefont{Punjabi}} \bibnamefont{et~al.},
  \bibinfo{journal}{Phys. Rev. C} \textbf{\bibinfo{volume}{71}},
  \bibinfo{pages}{055202} (\bibinfo{year}{2005}).

\bibitem[{\citenamefont{Gayou et~al.}(2001)}]{Gayou:2001qt}
\bibinfo{author}{\bibfnamefont{O.}~\bibnamefont{Gayou}} \bibnamefont{et~al.},
  \bibinfo{journal}{Phys. Rev. C} \textbf{\bibinfo{volume}{64}},
  \bibinfo{pages}{038202} (\bibinfo{year}{2001}).

\bibitem[{\citenamefont{Crawford et~al.}(2007)}]{Crawford:2006rz}
\bibinfo{author}{\bibfnamefont{C.~B.} \bibnamefont{Crawford}}
  \bibnamefont{et~al.}, \bibinfo{journal}{Phys. Rev. Lett.}
  \textbf{\bibinfo{volume}{98}}, \bibinfo{pages}{052301}
  (\bibinfo{year}{2007}).

\bibitem[{\citenamefont{Ron et~al.}(2007)}]{Ron:2007vr}
\bibinfo{author}{\bibfnamefont{G.}~\bibnamefont{Ron}} \bibnamefont{et~al.},
  \bibinfo{journal}{Phys. Rev. Lett.} \textbf{\bibinfo{volume}{99}},
  \bibinfo{pages}{202002} (\bibinfo{year}{2007}).

\bibitem[{\citenamefont{Lu et~al.}(1998)\citenamefont{Lu, Thomas, Tsushima,
  Williams, and Saito}}]{Lu:1997mu}
\bibinfo{author}{\bibfnamefont{D.-H.} \bibnamefont{Lu}},
  \bibinfo{author}{\bibfnamefont{A.~W.} \bibnamefont{Thomas}},
  \bibinfo{author}{\bibfnamefont{K.}~\bibnamefont{Tsushima}},
  \bibinfo{author}{\bibfnamefont{A.~G.} \bibnamefont{Williams}},
  \bibnamefont{and} \bibinfo{author}{\bibfnamefont{K.}~\bibnamefont{Saito}},
  \bibinfo{journal}{Phys. Lett. B} \textbf{\bibinfo{volume}{417}},
  \bibinfo{pages}{217} (\bibinfo{year}{1998}).

\bibitem[{\citenamefont{Saito et~al.}(2007)\citenamefont{Saito, Tsushima, and
  Thomas}}]{Saito:2005rv}
\bibinfo{author}{\bibfnamefont{K.}~\bibnamefont{Saito}},
  \bibinfo{author}{\bibfnamefont{K.}~\bibnamefont{Tsushima}}, \bibnamefont{and}
  \bibinfo{author}{\bibfnamefont{A.~W.} \bibnamefont{Thomas}},
  \bibinfo{journal}{Prog. Part. Nucl. Phys.} \textbf{\bibinfo{volume}{58}},
  \bibinfo{pages}{1} (\bibinfo{year}{2007}).

\bibitem[{\citenamefont{Meziani et~al.}(1992)}]{Meziani:1992xr}
\bibinfo{author}{\bibfnamefont{Z.~E.} \bibnamefont{Meziani}}
  \bibnamefont{et~al.}, \bibinfo{journal}{Phys. Rev. Lett.}
  \textbf{\bibinfo{volume}{69}}, \bibinfo{pages}{41} (\bibinfo{year}{1992}).

\bibitem[{E01()}]{E01-015}
\bibinfo{howpublished}{JLab E01-015, J. P. Chen, S. Choi and Z. E. Meziani,
  Spokespersons.}

\bibitem[{\citenamefont{Frank et~al.}(1996)\citenamefont{Frank, Jennings, and
  Miller}}]{Frank:1995pv}
\bibinfo{author}{\bibfnamefont{M.~R.} \bibnamefont{Frank}},
  \bibinfo{author}{\bibfnamefont{B.~K.} \bibnamefont{Jennings}},
  \bibnamefont{and} \bibinfo{author}{\bibfnamefont{G.~A.}
  \bibnamefont{Miller}}, \bibinfo{journal}{Phys. Rev. C}
  \textbf{\bibinfo{volume}{54}}, \bibinfo{pages}{920} (\bibinfo{year}{1996}).

\bibitem[{\citenamefont{Yakhshiev et~al.}(2003)\citenamefont{Yakhshiev,
  Meissner, and Wirzba}}]{Yakhshiev:2002sr}
\bibinfo{author}{\bibfnamefont{U.~T.} \bibnamefont{Yakhshiev}},
  \bibinfo{author}{\bibfnamefont{U.-G.} \bibnamefont{Meissner}},
  \bibnamefont{and} \bibinfo{author}{\bibfnamefont{A.}~\bibnamefont{Wirzba}},
  \bibinfo{journal}{Eur. Phys. J.} \textbf{\bibinfo{volume}{A16}},
  \bibinfo{pages}{569} (\bibinfo{year}{2003}).

\bibitem[{\citenamefont{Smith and Miller}(2004)}]{Smith:2004dn}
\bibinfo{author}{\bibfnamefont{J.~R.} \bibnamefont{Smith}} \bibnamefont{and}
  \bibinfo{author}{\bibfnamefont{G.~A.} \bibnamefont{Miller}},
  \bibinfo{journal}{Phys. Rev. C} \textbf{\bibinfo{volume}{70}},
  \bibinfo{pages}{065205} (\bibinfo{year}{2004}).

\bibitem[{\citenamefont{Horikawa and Bentz}(2005)}]{Horikawa:2005dh}
\bibinfo{author}{\bibfnamefont{T.}~\bibnamefont{Horikawa}} \bibnamefont{and}
  \bibinfo{author}{\bibfnamefont{W.}~\bibnamefont{Bentz}},
  \bibinfo{journal}{Nucl. Phys.} \textbf{\bibinfo{volume}{A762}},
  \bibinfo{pages}{102} (\bibinfo{year}{2005}).

\bibitem[{\citenamefont{Ciofi~degli Atti et~al.}(2007)\citenamefont{Ciofi~degli
  Atti, Frankfurt, Kaptari, and Strikman}}]{CiofidegliAtti:2007vx}
\bibinfo{author}{\bibfnamefont{C.}~\bibnamefont{Ciofi~degli Atti}},
  \bibinfo{author}{\bibfnamefont{L.~L.} \bibnamefont{Frankfurt}},
  \bibinfo{author}{\bibfnamefont{L.~P.} \bibnamefont{Kaptari}},
  \bibnamefont{and} \bibinfo{author}{\bibfnamefont{M.~I.}
  \bibnamefont{Strikman}}, \bibinfo{journal}{Phys. Rev. C}
  \textbf{\bibinfo{volume}{76}}, \bibinfo{pages}{055206}
  (\bibinfo{year}{2007}).

\bibitem[{\citenamefont{Schiavilla et~al.}(2005)\citenamefont{Schiavilla,
  Benhar, Kievsky, Marcucci, and Viviani}}]{Schiavilla:2004xa}
\bibinfo{author}{\bibfnamefont{R.}~\bibnamefont{Schiavilla}},
  \bibinfo{author}{\bibfnamefont{O.}~\bibnamefont{Benhar}},
  \bibinfo{author}{\bibfnamefont{A.}~\bibnamefont{Kievsky}},
  \bibinfo{author}{\bibfnamefont{L.~E.} \bibnamefont{Marcucci}},
  \bibnamefont{and} \bibinfo{author}{\bibfnamefont{M.}~\bibnamefont{Viviani}},
  \bibinfo{journal}{Phys. Rev. Lett.} \textbf{\bibinfo{volume}{94}},
  \bibinfo{pages}{072303} (\bibinfo{year}{2005}).

\bibitem[{\citenamefont{Seely et~al.}(2009)}]{Seely:2009gt}
\bibinfo{author}{\bibfnamefont{J.}~\bibnamefont{Seely}} \bibnamefont{et~al.},
  \bibinfo{journal}{Phys. Rev. Lett.} \textbf{\bibinfo{volume}{103}},
  \bibinfo{pages}{202301} (\bibinfo{year}{2009}).

\bibitem[{\citenamefont{Berz and Makino}(1997)}]{Berz:1997zz}
\bibinfo{author}{\bibfnamefont{M.}~\bibnamefont{Berz}} \bibnamefont{and}
  \bibinfo{author}{\bibfnamefont{K.}~\bibnamefont{Makino}}
  (\bibinfo{year}{1997}), \bibinfo{note}{17th IEEE Particle Accelerator
  Conference (PAC 97): Accelerator Science, Technology and Applications, 12-16
  May 1997, Vancouver, British Columbia, Canada}.

\bibitem[{\citenamefont{Besset et~al.}(1979)}]{Besset:1979sh}
\bibinfo{author}{\bibfnamefont{D.}~\bibnamefont{Besset}} \bibnamefont{et~al.},
  \bibinfo{journal}{Nucl. Instrum. Meth.} \textbf{\bibinfo{volume}{166}},
  \bibinfo{pages}{515} (\bibinfo{year}{1979}).

\bibitem[{\citenamefont{Afanasev et~al.}(2001)\citenamefont{Afanasev,
  Akushevich, and Merenkov}}]{PhysRevD.65.013006}
\bibinfo{author}{\bibfnamefont{A.~V.} \bibnamefont{Afanasev}},
  \bibinfo{author}{\bibfnamefont{I.}~\bibnamefont{Akushevich}},
  \bibnamefont{and} \bibinfo{author}{\bibfnamefont{N.~P.}
  \bibnamefont{Merenkov}}, \bibinfo{journal}{Phys. Rev. D}
  \textbf{\bibinfo{volume}{65}}, \bibinfo{pages}{013006}
  (\bibinfo{year}{2001}).

\bibitem[{\citenamefont{Blunden et~al.}(2005)\citenamefont{Blunden,
  Melnitchouk, and Tjon}}]{PhysRevC.72.034612}
\bibinfo{author}{\bibfnamefont{P.~G.} \bibnamefont{Blunden}},
  \bibinfo{author}{\bibfnamefont{W.}~\bibnamefont{Melnitchouk}},
  \bibnamefont{and} \bibinfo{author}{\bibfnamefont{J.~A.} \bibnamefont{Tjon}},
  \bibinfo{journal}{Phys. Rev. C} \textbf{\bibinfo{volume}{72}},
  \bibinfo{pages}{034612} (\bibinfo{year}{2005}).

\bibitem[{\citenamefont{Udias et~al.}(1999)\citenamefont{Udias, Caballero,
  Moya~de Guerra, Amaro, and Donnelly}}]{Udias:1999tm}
\bibinfo{author}{\bibfnamefont{J.~M.} \bibnamefont{Udias}},
  \bibinfo{author}{\bibfnamefont{J.~A.} \bibnamefont{Caballero}},
  \bibinfo{author}{\bibfnamefont{E.}~\bibnamefont{Moya~de Guerra}},
  \bibinfo{author}{\bibfnamefont{J.~E.} \bibnamefont{Amaro}}, \bibnamefont{and}
  \bibinfo{author}{\bibfnamefont{T.~W.} \bibnamefont{Donnelly}},
  \bibinfo{journal}{Phys. Rev. Lett.} \textbf{\bibinfo{volume}{83}},
  \bibinfo{pages}{5451} (\bibinfo{year}{1999}).

\bibitem[{\citenamefont{Udias and Vignote}(2000)}]{Udias:2000ig}
\bibinfo{author}{\bibfnamefont{J.~M.} \bibnamefont{Udias}} \bibnamefont{and}
  \bibinfo{author}{\bibfnamefont{J.~R.} \bibnamefont{Vignote}},
  \bibinfo{journal}{Phys. Rev. C} \textbf{\bibinfo{volume}{62}},
  \bibinfo{pages}{034302} (\bibinfo{year}{2000}).

\bibitem[{\citenamefont{Florizone et~al.}(1999)}]{Florizone:1999ba}
\bibinfo{author}{\bibfnamefont{R.~E.~J.} \bibnamefont{Florizone}}
  \bibnamefont{et~al.}, \bibinfo{journal}{Phys. Rev. Lett.}
  \textbf{\bibinfo{volume}{83}}, \bibinfo{pages}{2308} (\bibinfo{year}{1999}).

\bibitem[{\citenamefont{DeForest}(1983)}]{DeForest:1983vc}
\bibinfo{author}{\bibfnamefont{T.}~\bibnamefont{DeForest}},
  \bibinfo{journal}{Nucl. Phys.} \textbf{\bibinfo{volume}{A392}},
  \bibinfo{pages}{232} (\bibinfo{year}{1983}).

\bibitem[{\citenamefont{McNeil et~al.}(1983)\citenamefont{McNeil, Ray, and
  Wallace}}]{McNeil:1983yi}
\bibinfo{author}{\bibfnamefont{J.~A.} \bibnamefont{McNeil}},
  \bibinfo{author}{\bibfnamefont{L.}~\bibnamefont{Ray}}, \bibnamefont{and}
  \bibinfo{author}{\bibfnamefont{S.~J.} \bibnamefont{Wallace}},
  \bibinfo{journal}{Phys. Rev. C} \textbf{\bibinfo{volume}{27}},
  \bibinfo{pages}{2123} (\bibinfo{year}{1983}).

\bibitem[{\citenamefont{Horowitz}(1985)}]{Horowitz:1985tw}
\bibinfo{author}{\bibfnamefont{C.~J.} \bibnamefont{Horowitz}},
  \bibinfo{journal}{Phys. Rev. C} \textbf{\bibinfo{volume}{31}},
  \bibinfo{pages}{1340} (\bibinfo{year}{1985}).

\bibitem[{nop()}]{nopwia}
\bibinfo{howpublished}{Prior publications from E93-049 have reported the $R$
  ratio normalized to the PWIA. For simplicity in presentation, this text does
  not use that normalization.}

\bibitem[{\citenamefont{Dieterich et~al.}(2001)}]{Dieterich:2000mu}
\bibinfo{author}{\bibfnamefont{S.}~\bibnamefont{Dieterich}}
  \bibnamefont{et~al.}, \bibinfo{journal}{Phys. Lett. B}
  \textbf{\bibinfo{volume}{500}}, \bibinfo{pages}{47} (\bibinfo{year}{2001}).

\bibitem[{\citenamefont{Strauch et~al.}(2003)}]{Strauch:2002wu}
\bibinfo{author}{\bibfnamefont{S.}~\bibnamefont{Strauch}} \bibnamefont{et~al.}
  (\bibinfo{collaboration}{Jefferson Lab E93-049}), \bibinfo{journal}{Phys.
  Rev. Lett.} \textbf{\bibinfo{volume}{91}}, \bibinfo{pages}{052301}
  (\bibinfo{year}{2003}).

\end{thebibliography}

\end{document}